\documentclass[10pt,a4paper,twoside]{article}
\usepackage{epsfig}
\usepackage{baltlat6}
\usepackage{array}
\usepackage{here}
\begin{document}
\ \ \vspace{0.5mm} \setcounter{page}{1}

\titlehead{Baltic Astronomy, vol.\,24, 379--386, 2015}

\titleb{A STUDY OF DOUBLE- AND MULTI-MODE \\
RR LYRAE VARIABLES}

\begin{authorl}
\authorb{A. V. Khruslov}{1,2}
\end{authorl}

\begin{addressl}
\addressb{1}{Institute of Astronomy, Russian Academy of Sciences,
Pyatnitskaya str. 48, Moscow 119017, Russia; khruslov@bk.ru}
\addressb{2}{Sternberg Astronomical Institute, Moscow State
University, Universitetsky pr. 13, Moscow 119991, Russia.}
\end{addressl}

\submitb{Received: 2015 November 2; accepted: 2015 November 30}

\begin{summary}
We present the results of our new study of known RR Lyrae variable
stars. All observations available for these stars in the Catalina
Surveys were analyzed, and double-mode variations were identified.
We studied the Petersen diagram and period distribution for the
double-mode RR Lyrae variables in the Galactic field, pulsating in
the first-overtone and fundamental modes. The double-peaked
character of the period distribution was detected for Galactic
RR(B) stars, corresponding to Oosterhoff's classes of globular
clusters, which indicates that the age and evolution stage of
RR(B) stars in the field and RR~Lyrae variables in globular
clusters are probably the same. Besides, we discovered five
RRC~stars with two simultaneously excited non-radial pulsations
(equidistant triplets).
\end{summary}


\resthead{A study of double- and multi-mode RR Lyrae variables}
{A. V. Khruslov}

\sectionb{1}{INTRODUCTION}

For a long time, double periodicity was known for RR~Lyrae members
of globular star clusters. To our knowledge, the first star of
this type discovered in a globular cluster was V68 in M3
(Goranskij 1981). Smith (1995) listed the double-mode RR~Lyrae
variables known by that time in individual clusters. Jerzykiewicz
\& Wenzel (1977) discovered double-mode variations of AQ Leo, the
star that remained the only known double-mode RR~Lyrae star in the
Galactic field (with the exception of the triple-mode AC~And) for
more than a decade. The next two double-mode field RR~Lyrae stars,
EM~Dra and EN~Dra, were discovered by Clement et al. (1991).

For double-mode RR Lyrae stars, The General Catalogue of Variable
Stars (Samus et al. 2015) uses the designation RR(B).

Systematic discoveries and investigations of field RR(B) stars
were initiated by Cseresnjes (2001). In a photographic study of
fields in Sagittarius, he found double-mode variations of 53~RR
Lyrae stars, 40 of them believed to belong to the Sagittarius
dwarf galaxy and 13, to be stars of our Galaxy. Later, the ASAS-3
(Pojmanski 2002) and NSVS (Wozniak et al. 2004) automatic sky
surveys permitted the discoveries of several dozens of stars of
this type. Poleski (2014) found dozens of such stars using the
LINEAR data. By now, the OGLE-IV survey has identified 174
variables of this type in the region of the Galaxy's bulge
(Soszynski et al. 2014). The Catalog of periodic variable stars of
the Catalina Surveys (Drake et al. 2014) considers 502 variable
stars as double-mode RR Lyraes; it was later found that actually
only 165 of them belonged definitely to this type (Molnar et al.
2015); even among them, a small fraction are already known
double-mode stars.

Almost all currently known RR(B) variables pulsate in the
fundamental mode and first overtone, with the $P_1/P_0$ ratios
within $0.74-0.75$. Several stars with periods typical of RR~Lyrae
stars pulsate in the first and second overtones, $P_2/P_1\approx
0.80$, but, in our opinion, it is better to count these stars
among double-mode classical Cepheids: they are located near the
galactic plane and are very similar to double-mode Cepheids of the
Large and Small Magellanic Clouds in the same period range. We do
not know any field stars pulsating simultaneously in the first and
second overtones and possessing periods typical of RR Lyrae stars
at large distances from the galactic plane.

In most cases, the dominating mode is the first overtone. Most
often, the first overtone's pulsation amplitude is considerably
larger than that of the fundamental mode; cases of almost the same
amplitudes (but that of the first overtone being nevertheless
larger) are common enough; in still rarer cases, the amplitude of
the fundamental mode is slightly higher than that of the first
overtone; there are only very few RR(B) stars with the fundamental
mode amplitude significantly higher than that of the first
overtone.

Besides RR(B) stars with two radial pulsations, there are RR~Lyrae
stars with one or two additional non-radial pulsations. The
Blazhko effect in RRAB stars (fundamental-mode pulsators) can
usually be described with a superposition of an additional
non-radial mode. There exist a number of RRC stars (first-overtone
pulsators) that also have one or two simultaneously detected
non-radial modes. Some authors consider stars of this variability
type as RRC variables with the Blazhko effect.

RRC stars with a single non-radial pulsation whose frequency is
close to that of the first overtone were first discovered by Olech
et al. (1999) in the globular cluster M55 (three variables).
Alcock et al. (2000) identified 24 such variables in the Large
Magellanic Cloud (the MACHO project; the variability type
designated $RR1-\nu1$). A star of this kind, TYC 6556 00609 1, is
known in the galactic field (Antipin \& Jurcsik 2005). The
$P_2/P_1$ period ratios for stars of this type (assuming
$P_1>P_2$) are within $0.9 - 0.999$. The amplitude of the
secondary oscillation can be considerably lower than the amplitude
of the main one, but these amplitudes can also be virtually the
same.

RRC stars with two simultaneously excited non-radial pulsations
are often also called equidistant triplets: frequency differences
of the first and second non-radial modes with the main oscillation
are the same, i.e. the frequency of one of the non-radial
oscillations is higher than the $f_1$ frequency by some amount
$m$, while the frequency of another one is lower than the $f_1$
frequency by the same amount $m$. Let us denote all the
frequencies $f_1$, $f_1+m$, and $f_1-m$. Here the difference
between adjacent frequencies is not large, like in the case of the
RRC stars with a single non-radial pulsation described above.

Alcock et al. (2000) identified 28 variables of this type in the
Large Magellanic Cloud (the MACHO project, the variability type
designated as $RR1-BL$). Among stars of the Galactic field, the
case of NSV~07340 = V1141~Her is known (Antipin et al. 2010, $P_1
= 0.317152$~d, $m=0.030\ {\rm d}^{-1}$). Jurcsik et al. (2015)
found several RRC stars with non-radial pulsations in the globular
cluster M3, among them the equidistant-triplet star V140 ($P_1 =
0.33316$~d, $m=0.0689 {\rm d}^{-1}$).

\sectionb{2}{DOUBLE-MODE RR LYRAE STARS, RR(B) TYPE}

Since 2007, I have performed a search for double-mode RR Lyrae
variable stars using several available photometric surveys. I
detected (in some cases, with co-authors) double-mode variability
for 235 stars belonging to the RR(B) type, which currently
represent about one third of all known RR(B) stars in the Galactic
field. Most of them were discovered during the recent two years,
in the course of the analysis of the Catalina Surveys data (Drake
et al. 2009). Using these data, I mainly check the RRC stars with
considerable scatter of data points on the light curve, among
previously known as well as recently discovered stars from the
Catalina Surveys periodic variable star catalog (Drake et al.
2014). The limiting magnitude in minimum brightness for the
studied RR(B) stars is $19.4$ mag in the Catalina Surveys
photometric system ($CV$). The results of our search for RR(B)
stars are presented in a series of papers in {\itshape Peremennye
Zvezdy/Variable Stars}. The majority of the stars (207 variables)
were announced in three papers (Khruslov 2014, 2015ab). The first
of these papers contains also references to earlier publications.

Among the double-mode RR Lyrae variable stars we identified, there
are two interesting cases of variables with extreme periods. The
star USNO-B1.0 0822-0766869 (Khruslov 2015b, No. 103) has the
longest period among RR(B) variables, $P_0 = 0.600176$~d. The
light curve of USNO-B1.0 0822-0766869 is displayed in Fig.~1. A
slightly longer period ($P_0 = 0.60371$~d) is known only for the
star vs3f773 from Cseresnjes (2001), considered in the cited paper
to belong to stars of the Sagittarius dwarf galaxy. The variable
USNO-B1.0 0773-0284953 (Khruslov 2015b, No. 33) has almost the
shortest fundamental-mode period among field stars at sufficiently
large distances from the Galactic plane ($b = +37.4^\circ$; $P_0 =
0.447456$~d). Shorter-period RR(B) stars are known only in the
direction of the Galactic center: several dozens of stars from the
OGLE-III project and the star vs7f54 (Cseresnjes 2001) with $P_0 =
0.43574$~d, believed by Cseresnjes to be a variable in the Galaxy.

Besides, of interest is the rare case of USNO-B1.0 1344-0191047
(Khruslov 2014, No. 1) whose fundamental-mode amplitude is much
larger than that of the first overtone. The period ratio for this
star, $P_1/P_0 = 0.7460$, exceeds considerably that typical of the
corresponding fundamental-mode period ($P_0 = 0.482702$~d),
therefore, the star in the Petersen diagram lies much higher than
other stars with similar periods.

The Petersen diagram (relating the $P_1/P_0$ period  ratio and the
logarithm of the fundamental-mode period) for all known Galactic
field RR(B) F/1O variables (with the exception of the majority of
stars towards the Galactic center -- the Galactic bulge RR~Lyrae
stars from the OGLE-IV and MACHO projects and the Sagittarius
dwarf galaxy stars) is displayed in Fig. 2.

Based on our results and all other information available for RR(B)
stars (we did not take into account only stars towards the
Galactic center from surveys like OGLE), we plotted the period
distribution for the Galactic-field double-mode RR~Lyrae stars.
Data on 460 stars were used. The distribution, displayed in
Fig.~3, is double-peaked, with the primary maximum near the period
$P_0 = 0.48$~d and the secondary one near $P_0 = 0.54$~d.



\begin{figure}[!tH]
\vbox{
\centerline{\psfig{figure=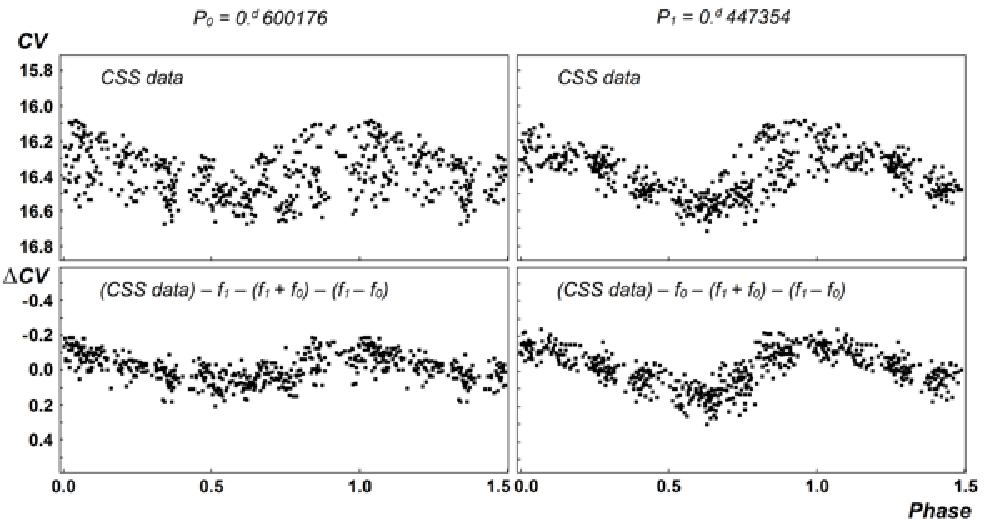,width=127mm,angle=0,clip=}}
\vspace{1mm} \captionb{1} {The light curves of USNO-B1.0
0822-0766869 (Khruslov 2015b, No. 103). Upper panels: raw data;
lower panels: the folded light curves with the other oscillation
pre-whitened.} }
\end{figure}


\begin{figure}[!tH]
\vbox{
\centerline{\psfig{figure=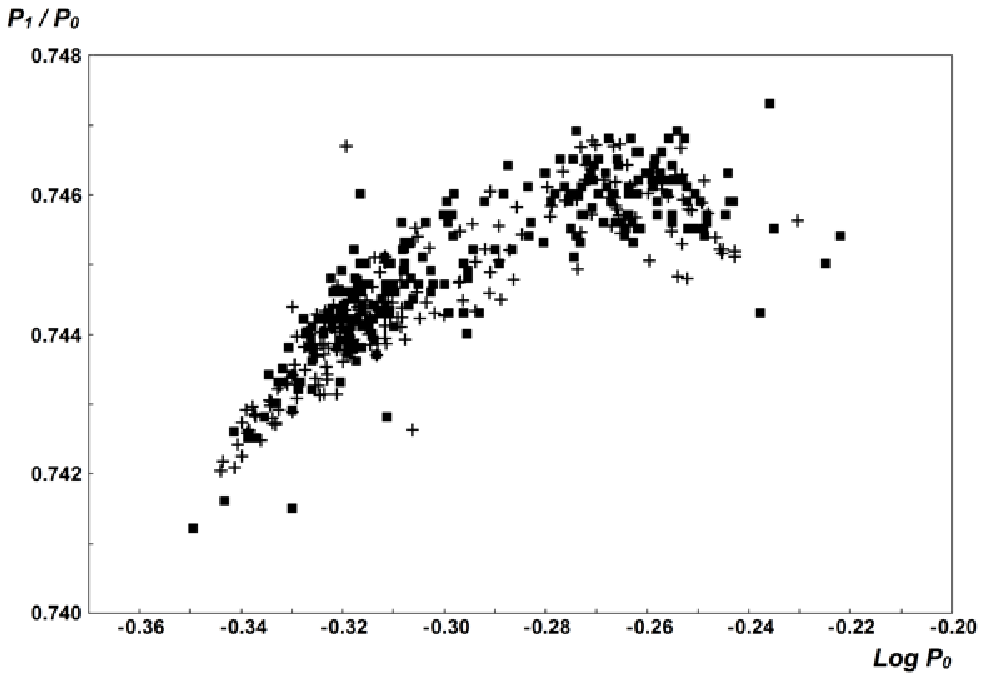,width=127mm,angle=0,clip=}}
\vspace{1mm} \captionb{2} {The Petersen diagram for the Galactic
field double-mode RR Lyrae F/1O variables. The squares represent
the F/1O RR(B) stars identified by us and the crosses denote
previously known RR(B) stars (according to
http://www.aavso.org/vsx/).} }
\end{figure}

\newpage
Earlier, the double-peaked character of the period distribution
for Galactic RR(B) stars was not obvious, the amount of data for
variables of this type being insufficient. A similar distribution
for the Large Magellanic Cloud does not show a double-peaked
character; is is marginally two-peaked for the Small Magellanic
Cloud (Soszynski et al. 2010).


\begin{figure}[!tH]
\vbox{
\centerline{\psfig{figure=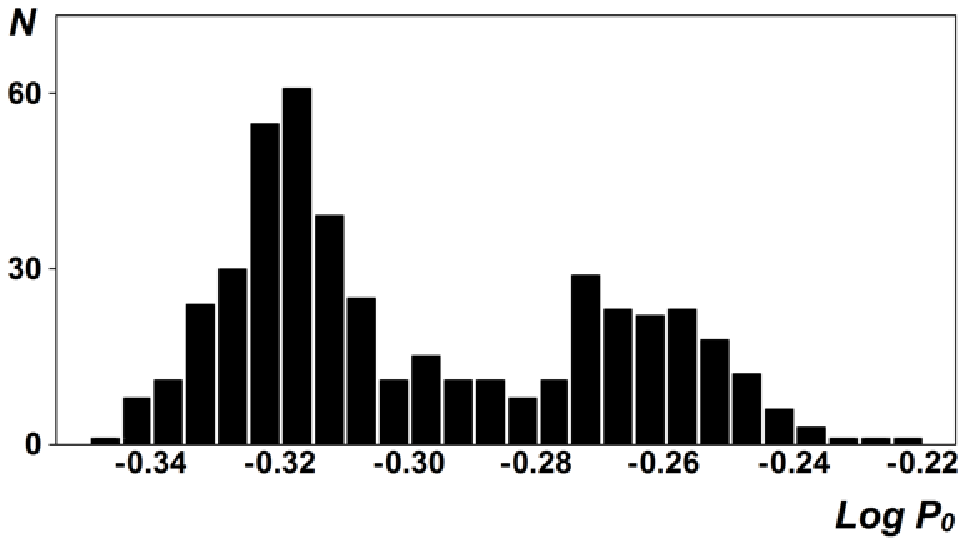,width=80mm,angle=0,clip=}}
\vspace{1mm} \captionb{3} {The period distribution of RR(B)
stars.} }
\end{figure}

The two maxima of the period distribution for RR(B) stars probably
coincide with maxima of period distributions for RR(B) stars in
globular clusters of Oosterhoff's classes I and II. Smith (1995),
with a reference to Clement et al. (1993), presents the Petersen
diagram with identified groups of RR(B) stars of the two
Oosterhoff's classes. Double-mode stars in the cluster of class~I
(IC~4499) have periods close to that of the main maximum of the
distribution for field RR(B) stars ($P_0 = 0.48$~d), while stars
in class~II clusters (M15 and M68) have periods close to that of
the distribution's secondary maximum ($P_0 = 0.54{-}0.55$~d).

If we remember that the distribution of periods for field RR~Lyrae
stars does not reveal peaks or gaps corresponding to Oosterhoff's
classes (Kukarkin 1975), probably due to the presence, in the
Galactic field, of RR~Lyrae stars with different element
abundances and ages, we can assume that the detected double-peaked
distribution indicates that age and evolutionary stage of RR(B)
stars in the field and RR~Lyrae stars in globular clusters are the
same.

\sectionb{3}{RRC STARS WITH NON-RADIAL PULSATION}

In the search for RR~Lyrae variables with two radial pulsations,
F/1O, I detected 16 RRC stars with a single excited non-radial
mode, its frequency close to that of the first-overtone radial
mode (Khruslov 2012, 2015c).

We earlier announced discoveries of equidistant triplets for two
RRC stars: TYC 3877 02198 1 (Khruslov 2010) and GSC 2010-00224
(Khruslov 2012). Later, we analyzed SuperWASP data (Butters et al.
2010) for TYC 3877 02198 1 and were not able to confirm the
presence of the third frequency.

Now, we announce five more cases of RRC stars with two
simultaneously excited non-radial pulsations (equidistant
triplets). All these variables were previously known as RRC stars
(with the exception of No.~2; Drake et al. 2014 earlier announced
it as an RR(B) star). We analyzed all observations available for
these stars in the Catalina Surveys online public archives using
the period-search software developed by Dr. V.P.~Goranskij for
Windows environment.

Information on the studied stars is presented in Tables 1 and 2.
Table 1 contains the equatorial coordinates (J2000), the star
number in the USNO-B1.0 catalog and the magnitudes at maximum and
minimum in the Catalina photometric system. Table 2 contains the
light elements and amplitudes: first-overtone period $P_1$;
frequency difference $m$ and periods of the non-radial pulsations
(frequencies $f_1+m$ and $f_1-m$); the first-overtone and
non-radial-mode epochs of maxima
(Epoch$_1$/Epoch$_{1+m}$/Epoch$_{1-m}$; 2455000 was subtracted
from all Julian dates); semi-amplitudes of the first-overtone and
non-radial-mode oscillations ($A_1/A_{1+m}/A_{1-m}$).

The light curves of the 5 new equidistant-triplet stars are
displayed in Fig.~4. Each row begins with the light curve for the
period $P_1$ (from initial data); then follow the light curves for
each of the three oscillations, the two other frequencies
excluded.

The periods of dominant oscillations (first overtone, $P_1$) for
all our equidistant triplets are within $0.280{-}0.363$ d. As for
amplitudes of the non-radial modes, the $A_{1+m}>A_{1-m}$ and
$A_{1+m}<A_{1-m}$ cases are equally frequent.


\begin{figure}[!tH]
\vbox{
\centerline{\psfig{figure=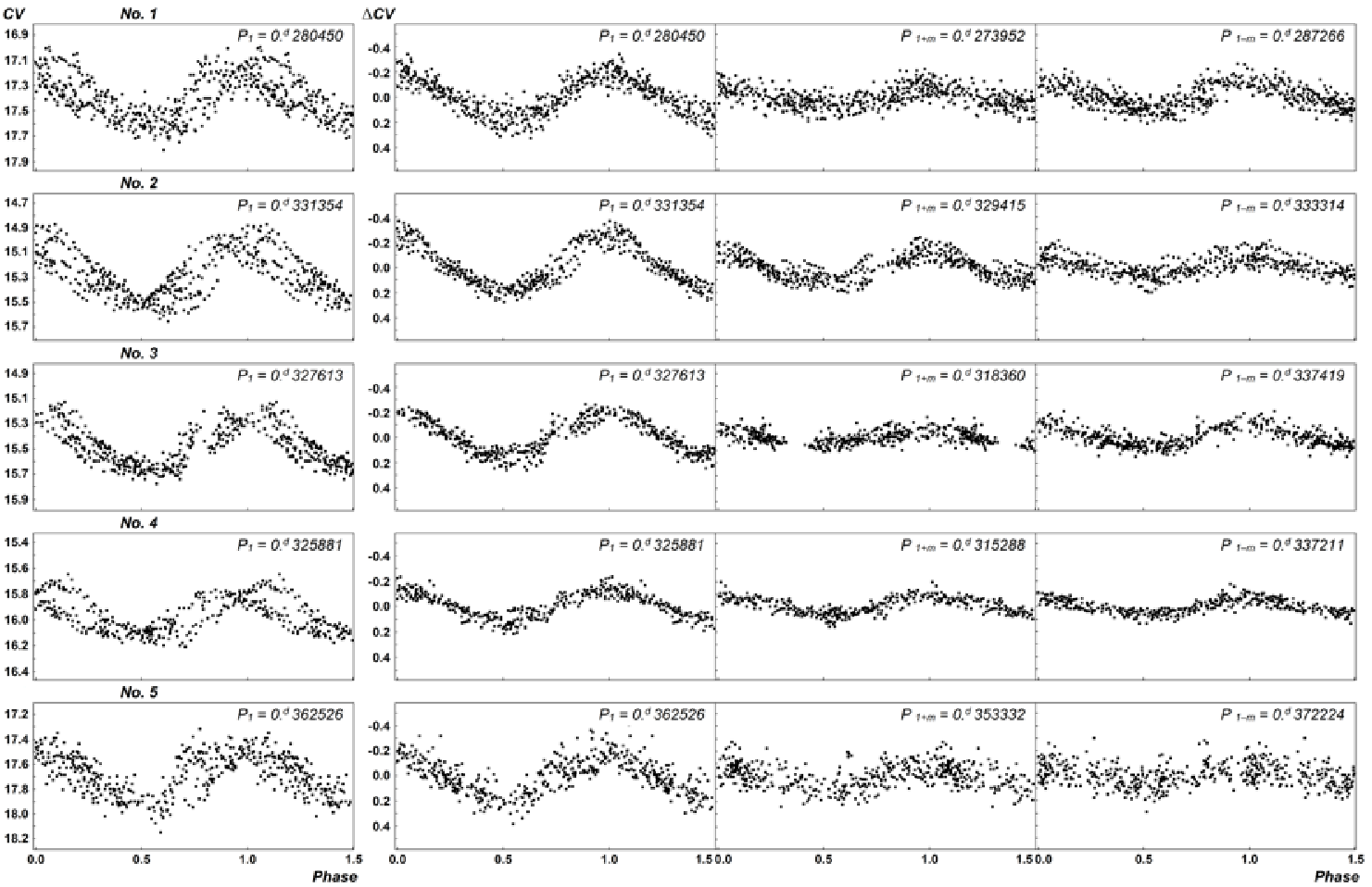,width=120mm,angle=0,clip=}}
\vspace{1mm} \captionb{4} {The light curves of RRC stars with two
non-radial pulsations (equidistant triplets). Left-hand column of
panels: raw data with the first overtone period; three right-hand
columns of panels: the folded light curves with the other
oscillations pre-whitened.} }
\end{figure}

\begin{table}[!t]
\begin{center}
\vbox{\footnotesize\tabcolsep=3pt
\parbox[c]{124mm}{\baselineskip=10pt
{\smallbf\ \ Table 1.}{\small\ Equidistant triplets: positions and
magnitudes.\lstrut}}
\begin{tabular}{r|c|c|c}
\hline
No. & Coordinates (2000.0) & USNO-B1.0 & mag (CV) \\

\hline

1 & 03$^h$09$^m$49$^s$.30 +28$^{\circ}$34$^{\prime}$05$^{\prime \prime}$.0 & 1185-0037622 & 17.00 $-$ 17.75 \\
2 & 11 39 37.27 $-$10 44 27.8 & 0792-0227178 & 14.87 $-$ 15.66 \\
3 & 13 06 10.90 $-$14 10 31.5 & 0758-0272311 & 15.13 $-$ 15.78 \\
4 & 16 20 25.94 +00 38 20.3 & 0906-0261394 & 15.65 $-$ 16.21 \\
5 & 16 51 26.51 +12 51 56.8 & 1028-0341556 & 17.32 $-$ 18.15 \\

\hline
\end{tabular}
}
\end{center}
\vskip-8mm
\bigskip
\end{table}

\begin{table}[!t]
\begin{center}
\vbox{\footnotesize\tabcolsep=3pt
\parbox[c]{124mm}{\baselineskip=10pt
{\smallbf\ \ Table 2.}{\small\ Equidistant triplets: light
elements and amplitudes\lstrut}}
\begin{tabular}{r|l|l|l|l|c|c}
\hline
No. & $P_1$, d & m, d$^{-1}$ & $P_{(1+m)}$, d & $P_{(1-m)}$, d & Epoch & $A$ \\

\hline

1 & 0.280450 & 0.0846 & 0.273952 & 0.287266 & 0.132 / 0.050 / 0.055 & 0.172 / 0.064 / 0.100  \\
2 & 0.331354 & 0.01776 & 0.329415 & 0.333314 & 0.216 / 0.128 / 0.110 & 0.200 / 0.097 / 0.068  \\
3 & 0.327613 & 0.0887 & 0.318360 & 0.337419 & 0.328 / 0.084 / 0.077 & 0.157 / 0.043 / 0.080  \\
4 & 0.325881 & 0.1031 & 0.315288 & 0.337211 & 0.162 / 0.013 / 0.113 & 0.118 / 0.067 / 0.056  \\
5 & 0.362526 & 0.0718 & 0.353332 & 0.372224 & 0.325 / 0.107 / 0.340 & 0.151 / 0.070 / 0.056  \\

\hline
\end{tabular}
}
\end{center}
\vskip-8mm
\bigskip
\end{table}

\sectionb{4}{CONCLUSIONS}

Our new study of known RR Lyrae variable stars from the Catalina
Surveys data that has revealed a number of new double-mode
variables increases considerably the material available for future
statistical analysis. The two-peaked character of the period
distribution was detected for Galactic RR(B) stars, corresponding
to Oosterhoff's classes of globular clusters. The five new cases
of equidistant triplets are among the first ones identified in the
Galaxy's field. We found that the phased light curves of
double-mode stars, plotted with the first-overtone period, have a
typical shape and scatter, therefore it is possible to preliminary
identify (prior to the frequency analysis) equidistant-triplet
stars among other RRC~variables.

\thanks{The author is grateful to Dr. V.P.~Goranskij for providing a software for the
light-curve analysis. Thanks are also due to Drs. S.V.~Antipin and
N.N.~Samus for helpful discussions. This study was supported by
the Russian Foundation for Basic Research (grant 13-02-00664) and
the Programme ``Transitional and Outburst Processes in the
Universe'' of the Presidium of the Russian Academy of Sciences.}

\References

\refb Alcock C., Allsman R., Alves D. R., et al. 2000, ApJ, 542,
257

\refb Antipin S. V., Jurcsik, J. 2005, IBVS, No. 5632

\refb Antipin S. V., Kazarovets E. V., Pastukhova E. N. 2010,
Perem. Zvezdy/Variable Stars Suppl., 10, No. 33

\refb Butters O. W., West R. G., Anderson D. R., et al. 2010,
A\&A, 520, L10

\refb Clement C. M., Kinman T. D., Suntzeff N. B. 1991, ApJ, 372,
273

\refb Clement C. M., Ferance S., Simon N. R. 1993, ApJ, 412, 183

\refb Cseresnjes P. 2001, A\&A, 375, 909

\refb Drake A. J., Djorgovski S. G., Mahabal A. et al. 2009, ApJ,
696, 870

\refb Drake A. J., Graham M. J., Djorgovski S. G. et al. 2014,
ApJS., 213, 9

\refb Goranskij V. P. 1981, IBVS, No. 2007

\refb Jerzykiewicz M., Wenzel W. 1977, Acta Astron., 27, 35

\refb Jurcsik J., Smitola P., Hajdu G. et al. 2015, ApJS, 219, 25

\refb Khruslov A. V. 2010, Perem. Zvezdy/Variable Stars Suppl.,
10, No. 32

\refb Khruslov A. V. 2012, Perem. Zvezdy/Variable Stars Suppl.,
12, No. 18

\refb Khruslov A. V. 2014, Perem. Zvezdy/Variable Stars, 34, No. 3

\refb Khruslov A. V. 2015a, Perem. Zvezdy/Variable Stars, 35, No.
1

\refb Khruslov A. V. 2015b, Perem. Zvezdy/Variable Stars, 35, No.
4

\refb Khruslov A. V. 2015c, Perem. Zvezdy/Variable Stars, 35, No.
5

\refb Kukarkin B. V. 1975 in ``Variable Stars and Stellar
Evolution'' (IAU Symp. No. 67), eds. V. E. Sherwood, L. Plaut,
Dordrecht and Boston: D.~Reidel Publishing Co., p. 511

\refb Molnar L., Szabo R., Moskalik P. A. et al. 2015, MNRAS, 452,
4283

\refb Olech A., Kaluzny J., Thompson I.B. et al. 1999, AJ, 118,
442

\refb Pojmanski G. 2002, Acta Astron., 52, 397

\refb Poleski R. 2014, PASP, 126, 509

\refb Samus N. N., Durlevich O. V., Kazarovets E. V. et al. 2015,
General Catalogue of Variable Stars, Centre de Donnees
Astronomiques de Strasbourg, B/gcvs, Moscow online version:
http://www.sai.msu.su/gcvs/gcvs/

\refb Smith H. A. 1995, ``RR Lyrae stars''. Cambridge Astrophysics
Series, V.~27, Cambridge, UK: Cambridge University Press

\refb Soszynski I., Udalski A., Szymanski M.K., et al. 2010, Acta
Astron., 60, 165

\refb Soszynski I., Udalski A., Szymanski M.K. 2014, Acta Astron.,
64, 1

\refb Wozniak P. R., Vestrand W. T., Akerlof C. W. et al. 2004,
AJ, 127, 2436

\end{document}